\documentclass[twocolumn,showpacs,preprintnumbers]{revtex4}%
\usepackage{amssymb}
\usepackage{amsfonts}
\usepackage{amsmath}
\usepackage{graphicx}
\usepackage{dcolumn}
\usepackage{bm}%
\begin{document}
\preprint{APS/123-QED}
\title{Nonlinear Coupling in Nb/NbN Superconducting Microwave Resonators}
\author{B. Abdo}
\email{baleegh@tx.technion.ac.il}
\author{E. Segev}
\author{O. Shtempluck}
\author{E. Buks}
\affiliation{Microelectronics Research Center, Department of Electrical Engineering,
Technion, Haifa 32000, Israel}
\date{\today}

\begin{abstract}
In this experimental study we show that the coupling between Nb/NbN
superconducting microwave resonators and their feedline can be made amplitude
dependent. We employ this mechanism to tune the resonators into critical
coupling condition, a preferable mode of operation for a wide range of
applications. Moreover we examine the dependence of critical coupling state on
other parameters such as temperature and magnetic field. Possible novel
applications based on such nonlinear coupling are briefly discussed.

\end{abstract}
\maketitle




Recently, there has been a growing scientific interest in nonlinear
directional couplers especially in the optical regime \cite{squeezing}. It has
been shown in a series of publications, that operating such devices by means
of parametric downconversion, may exhibit some important quantum phenomena,
such as generation of quadrature-squeezed states \cite{squeezing}, occurrence
of quantum Zeno and anti-Zeno effects \cite{zeno}, non-classical correlations
\cite{EPR}, and quantum entanglement \cite{entanglement}. Such effects, if
generated and engineered, would have immediate and significant implications on
a wide range of fields, ranging from basic science to optical communication
systems, and quantum computation \cite{correlated},\cite{statics}. The
dependence of these mentioned effects on optimal coupling strengths and cavity
detunings are almost evident \cite{squeezing},\cite{EPR}.

In this study we investigate the coupling between nonlinear NbN/Nb microwave
resonators and their feedlines, as a function of injected power, ambient
temperature and applied magnetic field. We show that the coupling in these
resonators could be, to a large extent, dependent on externally applied
parameters, allowing us thus to tune the resonators into critical coupling
condition, which is a coupling state where the power transferred to the device
is maximized and no power reflection is present. Such a state occurs as both
impedances of the device and of the feedline match.

Theoretically critical coupling can be achieved by either tuning the coupling
parameters or by altering the internal/external losses of the device. Such
control was reported in optical systems \cite{Control SOA}, where critical
coupling was obtained in microring resonator by altering the internal loss
inside the ring by means of integrated semiconductor optical amplifiers. Other
critical coupling achieved in a fiber ring resonator configuration may be used
in optical communication systems i.e. switches and modulators \cite{cc
Yariv},\cite{Control of cc ring-fiber conf}.

\begin{figure}
[ptb]
\begin{center}
\includegraphics[
height=2.4128in,
width=2.9827in
]%
{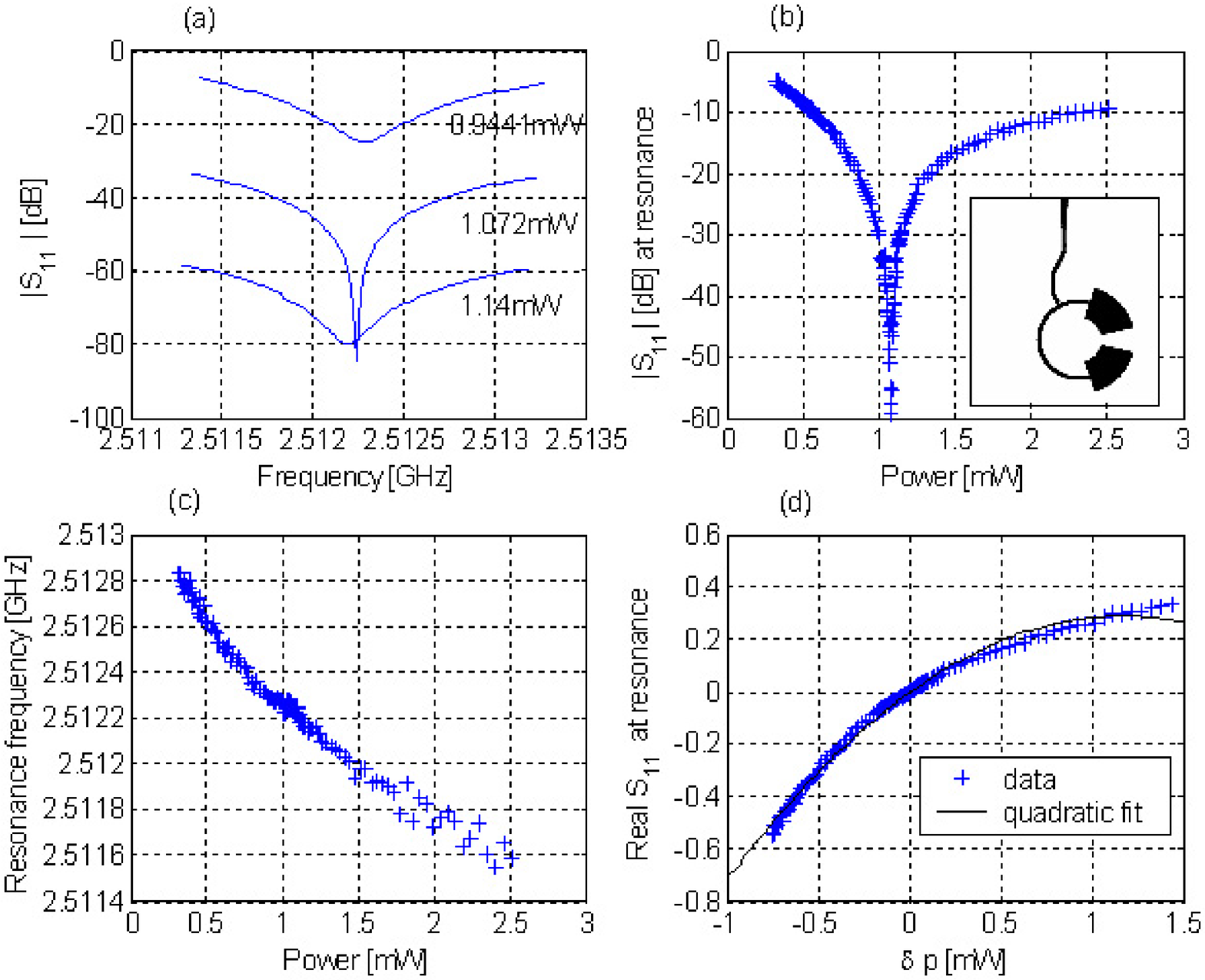}%
\caption{Data analysis of NbN \ resonator first mode. (a) $\left\vert
S_{11\text{ }}\right\vert $ graphs representing $p<p_{c},$ $p=p_{c}$ and
$p>p_{c}$ cases. The resonance curves were shifted vertically by a constant
offset for clarity. (b) $\left\vert S_{11\text{ }}\right\vert $ minimum as a
function of input power. The minimum of the graph is at $p_{c}.$ (c) Resonance
frequency vs. input power$.$ (d) Real $S_{11}$ at resonance vs. power
difference from critical coupling power. The resonator geometry is shown in
the inset of subplot (b).}%
\label{B2_1_ana}%
\end{center}
\end{figure}

\begin{figure}
[ptb]
\begin{center}
\includegraphics[
height=2.3912in,
width=3.1445in
]%
{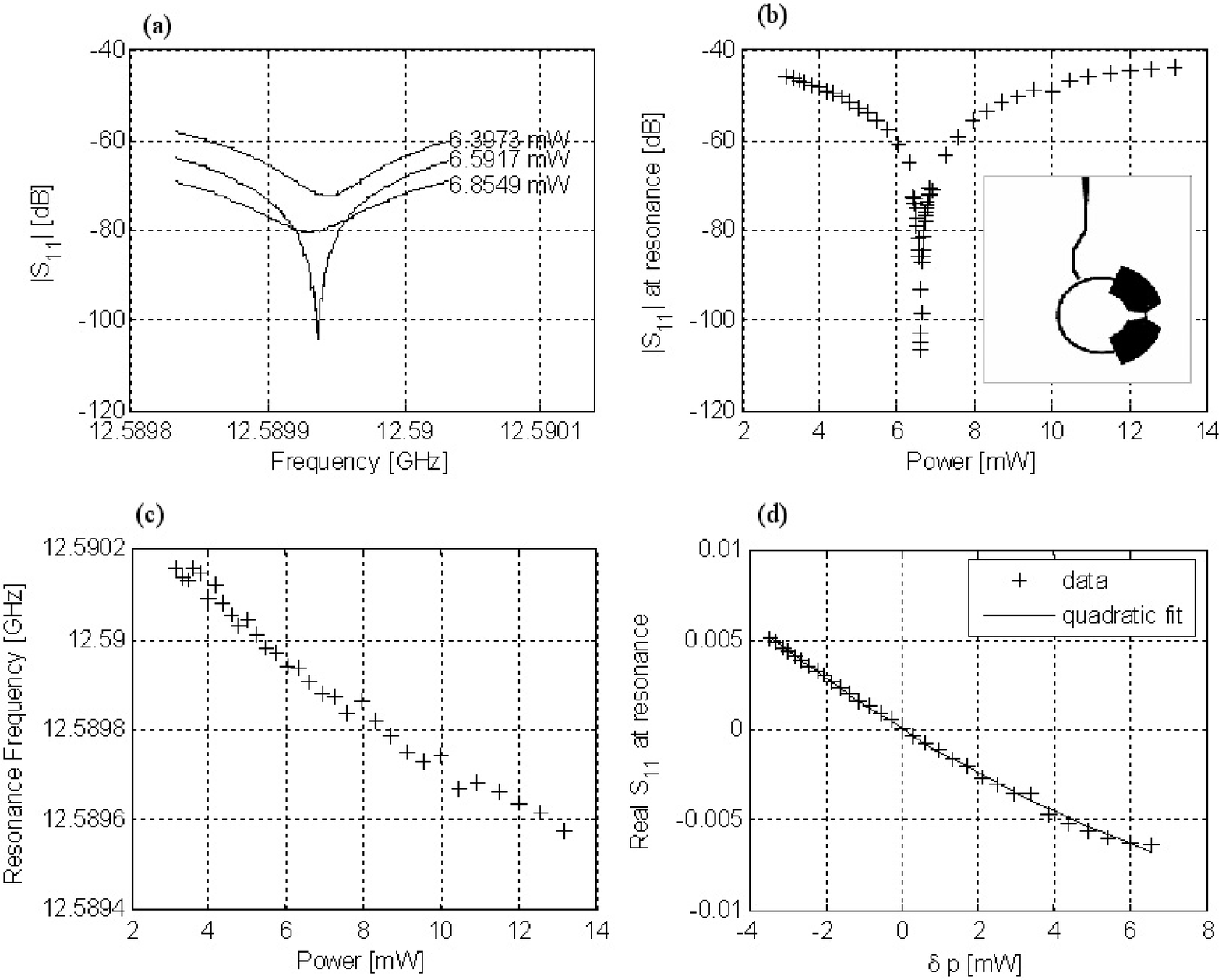}%
\caption{Data analysis of Nb resonator critical coupling at
$12.589\mathrm{GHz}$ resonance. (a) $\left\vert S_{11\text{ }%
}\right\vert $ graphs representing $p<p_{c},$ $p=p_{c}$ and $p>p_{c}$ cases.
The resonance curves were shifted vertically by a constant offset for clarity.
(b) $\left\vert S_{11\text{ }}\right\vert $ minimum as a function of input
power. The minimum of the graph is at $p_{c}$ . (c) Resonance frequency vs.
input power$.$ (d) Real $S_{11}$ at resonance vs. power difference $\delta p$
from critical coupling power. The resonator geometry is shown in the inset of
subplot (b).}%
\label{Nb_ana}%
\end{center}
\end{figure}

Our fabricated resonators were assembled in a standard stripline geometry, and
were housed in a gold plated Faraday package made of Oxygen Free High
Conductivity (OFHC) Cupper. The layout of the Nb and NbN resonators are
presented in the insets of Figs.\ref{Nb_ana} and \ref{B2_1_ana} respectively.
The resonators were dc-magnetron sputtered, at room temperature, on sapphire
substrates with dimensions 34$%
\mathrm{mm}%
$ X 30$%
\mathrm{mm}%
$ X 1$%
\mathrm{mm}%
$ each. A gap of 0.5 mm was set between the feedline and the resonators. The
thickness of the Nb sputtered film is $2200%
\mathrm{\text{\AA}}%
,$ whereas the thickness of the NbN film is $3000%
\mathrm{\text{\AA}}%
$. Following the deposition, the Nb resonator was patterned using standard
photolithography process and dry etched with photo resist mask. The sputtering
parameters of the NbN superconducting microwave resonator have been discussed
elsewhere \cite{nonlinear features BB}.

\begin{figure}
[ptb]
\begin{center}
\includegraphics[
height=2.0591in,
width=2.6039in
]%
{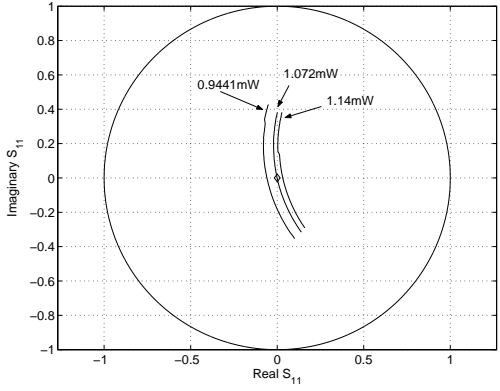}%
\caption{ NbN resonator first mode $S_{11}\left(  \omega\right)  $ curve,
drawn in the complex plane for three input powers $0.944\mathrm{mW}$,
$1.072$ $\mathrm{mW}$, $1.14\mathrm{mW},$ with
$2\mathrm{MHz}$ span. The resonator changes from overcoupled to
undercoupled state as the power is increased through $p_{c}=1.072$
$\mathrm{mW}.$ The unit circle and the origin are also plotted on the
same axis for visual comparison.}%
\label{NbN_circ}%
\end{center}
\end{figure}

The measurements of the resonators were done using the \textit{S} parameter of
a vector network analyzer, while varying the input power or the ambient
temperature in small steps. In Figs. \ref{NbN_circ},\ref{B2_1_ana} we show a
power induced critical coupling data obtained at the first mode of the NbN
resonator. In Fig. \ref{NbN_circ} we plot the $S_{11}\left(  \omega\right)  $
curves of the first mode in the complex plane, corresponding to three input
powers $0.944%
\mathrm{mW}%
$, $1.072$ $%
\mathrm{mW}%
$, $1.14%
\mathrm{mW}%
$ in the vicinity of the resonance, representing $p<p_{c},$ $p=p_{c},$
$p>p_{c}$ states respectively, where $p_{c}$ is the injected power level at
which critical coupling is achieved. These three trajectories show clearly
that the NbN resonator at its first mode has transformed from overcoupled
(strongly coupled), to undercoupled state (loosely coupled) through critically
coupled condition via input power increase only. In Fig. \ref{B2_1_ana} (a) we
show the resonance curves $\left\vert S_{11}\right\vert $ of the three input
powers, presented previously in Fig. \ref{NbN_circ}, in the vicinity of the
resonance. The resonances were shifted vertically for clarity. In Fig.
\ref{B2_1_ana} (b) we present the nonlinear evolution of the $\left\vert
S_{11}\right\vert $ minimum at resonance as the input power is increased
through the critical coupling power. The minimum of the graph $\sim$ $-60$ dB
at $p_{c}=1.072$ $%
\mathrm{mW}%
,$ corresponds to critical coupling condition where there is no power
reflection. The gradual shift of the resonance frequency towards lower
frequencies as the input power is increased, is plotted in Fig. \ref{B2_1_ana} (c).%

%

\begin{figure}
[ptb]
\begin{center}
\includegraphics[
height=1.2851in,
width=3.4783in
]%
{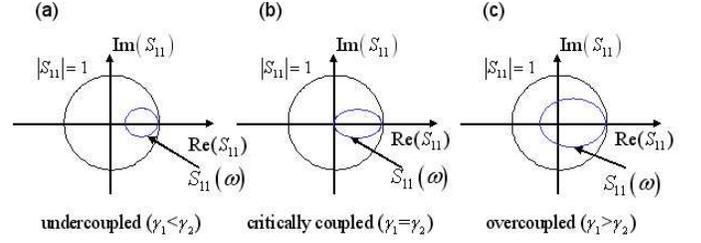}%
\caption{Resonator coupling states as defined by their $S_{11}(\omega)$ curves
in the vicinity of the resonance $\omega_{0},$ (see Eq.(\ref{S11}))
corresponding for (a) undercoupled state (b) critically coupled state (c)
overcoupled state.}%
\label{ccterm}%
\end{center}
\end{figure}

In order to identify and characterize the coupling mechanism responsible for
this dependence on the drive amplitude, we apply the following universal
expression for reflection amplitude of a linear resonator near resonance
\cite{Yurke Eyal}%

\begin{equation}
S_{11}(\omega)=\frac{i(\omega_{0}-\omega)-(\gamma_{1}-\gamma_{2})}%
{i(\omega_{0}-\omega)+(\gamma_{1}+\gamma_{2})}, \label{S11}%
\end{equation}

where $\gamma_{1}$ and $\gamma_{2}$ are real coupling factors associated with
the feedline-resonator and the resonator-reservoir (dissipation) couplings
respectively. However, unlike linear oscillators where $\gamma_{1}$,
$\gamma_{2}$ and $\omega_{0}$ are constants, here we assume that these factors
depend on some externally applied parameter $x,$ such as input power,
temperature, and magnetic field. This generalization could be justified by
noting that the $\left\vert S_{11}\left(  \omega\right)  \right\vert $ curves
measured in the vicinity of the resonance, are to a good approximation
symmetrical Lorentzians. The different possible coupling states of the
resonator as described generally by Eq.(\ref{S11}) are shown schematically in
Fig. \ref{ccterm}.%

By expanding $\gamma_{1}$ and $\gamma_{2}$ near critical coupling drive
$x_{c},$ to lowest order in $\delta x=x-x_{c},$ we get $\gamma_{1}%
=\gamma+g_{1}\delta x$ and $\gamma_{2}=\gamma+g_{2}\delta x,$ where
$\gamma,g_{1} $ and $g_{2}$ are real constants.

Substituting for $\gamma_{1}$ and $\gamma_{2}$ in the vicinity of $x_{c}$ in
Eq.(\ref{S11}) at the corresponding resonance $\omega=\omega_{0},$ yields%

\begin{equation}
S_{11}\left(  \omega_{0}\right)  =a_{1}\delta x+a_{2}\left(  \delta x\right)
^{2}+O\left(  \left(  \delta x\right)  ^{3}\right)  , \label{poly fit}%
\end{equation}

where $a_{1}=\left(  g_{2}-g_{1}\right)  /2\gamma,$ $a_{2}=\left(  g_{1}%
^{2}-g_{2}^{2}\right)  /4\gamma^{2}.$

In order to evaluate the critical coupling constant $\gamma$ characterizing
the resonance presented earlier, we applied a fit to the resonance curve at
$p_{c}$ using Eq.(\ref{S11}) and obtained $\gamma=1.3%
\mathrm{MHz}%
.$ Moreover, we extracted $g_{1}$ and $g_{2}$ by applying a quadratic
polynomial fit as in Eq.(\ref{poly fit}) to the experimental data of real
$S_{11}$ at resonance as a function of $\delta p.$ The data and the quadratic
fit are presented in Fig. \ref{B2_1_ana} (d), yielding the following nonlinear
coupling rates $g_{1}=-0.1%
\mathrm{MHz}%
/%
\mathrm{mW}%
$, $g_{2}=1.2%
\mathrm{MHz}%
/%
\mathrm{mW}%
.$ Having $\left\vert g_{2}\right\vert >\left\vert g_{1}\right\vert $ and
$g_{2}>0,$ implies that the critical coupling in this case, is caused mainly
by the increase of the dissipation coupling as the input power is increased.

In the Nb resonator, power induced critical coupling was attained at $12.589%
\mathrm{GHz}%
$ resonance as shown in Fig. \ref{Nb_ana}. The resonator changes from
undercoupled state to overcoupled state as the power is increased through the
critical coupling power at $6.59%
\mathrm{mW}%
$ . The coefficients $\gamma,g_{1},g_{2}$ extracted for this case are $48.5%
\mathrm{MHz}%
,$ $1.6%
\mathrm{MHz}%
/%
\mathrm{mW}%
,$ $1.5%
\mathrm{MHz}%
/%
\mathrm{mW}%
$ respectively. Implying surprisingly that what led to critical coupling in
this case, is the increase of the feedline-resonator coupling at higher rate
than the increase of the dissipation coupling rate, since we have $\left\vert
g_{1}\right\vert >\left\vert g_{2}\right\vert $, $g_{1}>0$. Such increase in
the feedline-resonator coupling could be attributed to a nonlinear
redistribution of the standing-wave voltage-mode inside the resonator relative
to the feedline position, leading to amplitude dependence of $\gamma_{1}.$%

Based on our experimental demonstration of nonlinear coupling between
superconducting stripline elements it is interesting to examine the
feasibility of implementing some of the theoretical proposals presented in
Ref. [1-6] in the microwave regime. \ For that end we consider a resonator
made of two stripline elements having nonlinear coupling factor $g$ between
them. \ The resonator is also coupled to a feed line employed for delivering
the input and output signals. \ Consider operating close to a resonance having
angular frequency $\omega_{0}$ and damping rate $\gamma$. \ A detailed
analysis of this model is beyond the scope of the present letter. \ However,
employing dimensionality analysis and some simplifying assumptions one finds
that the onset of nonlinear instability and bifurcation is expected in such a
system at input power level of order $p_{NL}^{in}\cong\gamma/g$. \ In the
regime of operation when the input power is close to $p_{NL}^{in}$ effects
such as intermodulation gain and quantum squeezing \cite{squeezing} are
expected to become noticeable \cite{Yurke Eyal}. \ Based on our results with
the Nb resonator, we estimate from the experimental values of $\gamma$ and
$g_{1}$ (assuming $g=g_{1}$) that such effects can be implemented with a
moderate input power of order $10^{-2}%
\mathrm{W}%
$. \ On the other hand, effects such as Zeno and anti-Zeno \cite{zeno} require
much stronger nonlinear coupling. \ We estimate that in the limit of zero
temperature such effects will become noticeable only when the input power
becomes comparable to $p_{\varphi}^{in}\cong\gamma/\hslash\omega_{0}g^{2}$.
\ Again, assuming the experimental values extracted from the data of the Nb
resonator, one finds $p_{\varphi}^{in}\cong10^{12}%
\mathrm{W}%
$ - far too high for any practical implementation. \ However, further analysis
is required to explore other possibilities of implementing such effects in the
microwave regime.%

\begin{figure}
[ptb]
\begin{center}
\includegraphics[
height=2.3999in,
width=3.2716in
]%
{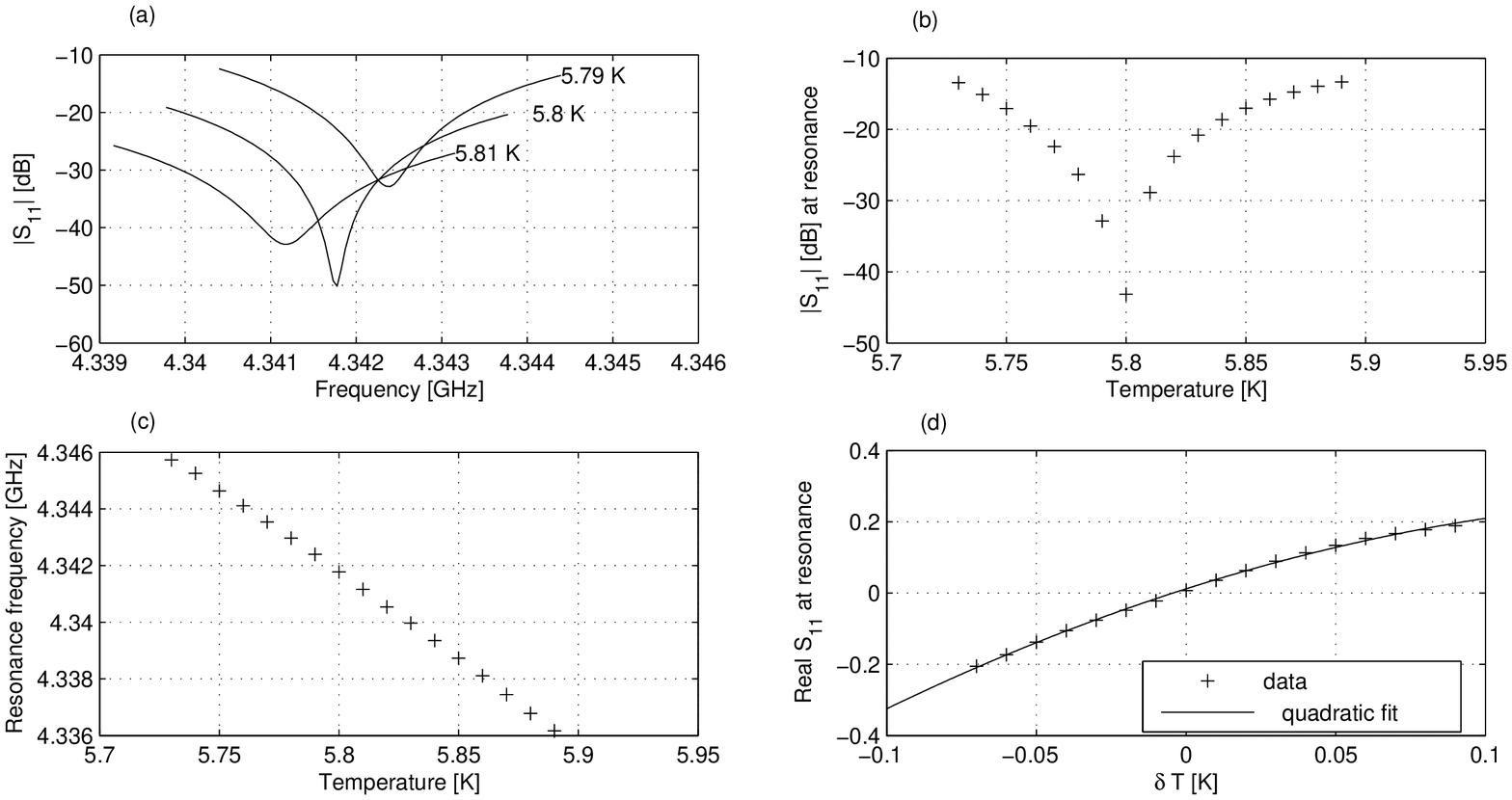}%
\caption{Data analysis of NbN second mode temperature induced critical
coupling. (a) $\left\vert S_{11\text{ }}\right\vert $ graphs representing
$T<T_{crit},T=T_{crit}$ and $T>T_{crit}$ cases. The resonance curves were
shifted vertically by a constant offset for clarity. (b) $\left\vert
S_{11\text{ }}\right\vert $ minimum as a function of temperature, the minimum
of the graph is at $T_{crit}$ . (c) Resonance frequency vs. temperature$.$ (d)
Real $S_{11}$ at resonance vs. temperature difference $\delta T$ from critical
coupling temperature. }%
\label{B2_2_ana_T}%
\end{center}
\end{figure}

In addition to input power tuning, we have succeeded to obtain critical
coupling using other modes of operation, one of which was tuning the ambient
temperature. To demonstrate this, we applied a constant input power to the NbN
resonator and measured its second resonance response as we increased the
ambient temperature in 0.01$%
\mathrm{K}%
$ steps. The input power applied was $0.1%
\mathrm{mW}%
$ , lower than $p_{c}$ of that resonance at 4.2 $%
\mathrm{K}%
,$ which was about $0.209$ $%
\mathrm{mW}%
$. The resonator changed from overcoupled state at $T<T_{crit}$ to
undercoupled state at $T>T_{crit}$. $T_{crit}$ was found to be 5.8$%
\mathrm{K}%
.$ The measurement results of this mode of operation are summarized in Fig.
\ref{B2_2_ana_T}. By applying similar fitting procedures as was explained
before one can obtain the following constants: $\gamma=4.2%
\mathrm{MHz}%
,g_{1}=-0.6%
\mathrm{MHz}%
/%
\mathrm{K}%
,g_{2}=220%
\mathrm{MHz}%
/%
\mathrm{K}%
,$ indicating that what led to critical coupling condition, in this case, was
the increase of the dissipation coupling as a result of the temperature rise.%

\begin{figure}
[ptb]
\begin{center}
\includegraphics[
height=1.9536in,
width=2.8037in
]%
{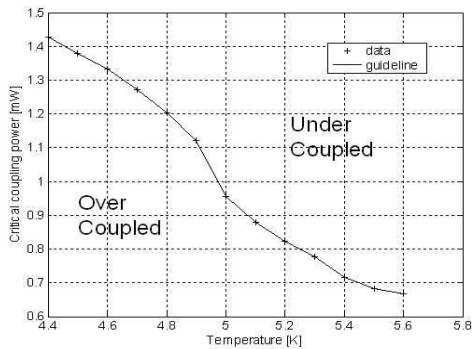}%
\caption{A plot displaying the critical coupling power dependence of the first
mode of the NbN resonator on ambient temperature. The critical coupling power
decreases as the temperature is increased. }%
\label{CC vs temp}%
\end{center}
\end{figure}

Moreover to better understand the dual relation between the input power and
the ambient temperature, additional measurement was carried out on the first
mode of the NbN resonator. We varied the temperature of the resonator as a
parameter and at each given temperature, we scanned the input power in search
for the critical coupling power $p_{c}$ . The measurement result is shown in
Fig. \ref{CC vs temp}, where we see that in the case of the first mode of the
NbN resonator at $\sim2.5%
\mathrm{GHz}%
$, increasing the ambient temperature decreased the input power at which
critical coupling occurred, and thus divided the $T-p^{in}$ plane into two
regions representing the overcoupled and undercoupled resonator states,
approximately separated by the guideline shown in the figure.%

\begin{figure}
[ptb]
\begin{center}
\includegraphics[
height=2.0401in,
width=3.0528in
]%
{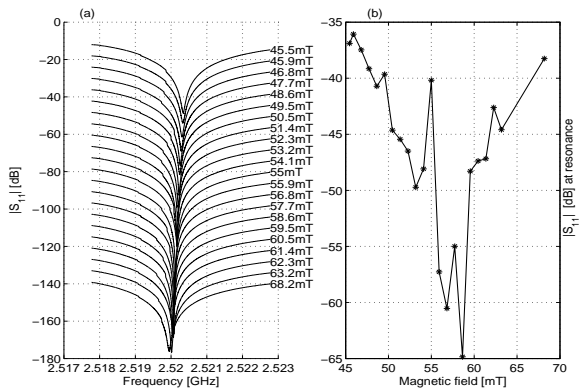}%
\caption{Critical coupling achieved by magnetic field at the first mode
resonance of the NbN resonator while applying constant input power of
$1.23\mathrm{mW}$. (a) $\left\vert S_{11}\right\vert $ vs. frequency
corresponding to an increased applied magnetic field. The resonance curves
were shifted vertically by a constant offset for clarity. (b) The $\left\vert
S_{11}\right\vert $ value at the resonance frequency. }%
\label{CC mag}%
\end{center}
\end{figure}

Critical coupling condition was also achieved in the NbN resonator by means of
applied magnetic field. To show this we set a constant input power of $1.23%
\mathrm{mW}%
$ (the critical coupling power at the time of the measurement in the absence
of magnetic field was $1.259%
\mathrm{mW}%
$), and increased a perpendicular magnetic field by small steps. The results
of this mode of operation, unlike power and temperature, showed a distinct
instability and frequent transitions around the critical coupling state as can
be inferred from Fig. \ref{CC mag}. A possible explanation to this instable
behavior could be vortex penetration into the grain boundaries of our NbN
film, as $H_{c_{1}}$ in NbN could be in the order of $35%
\mathrm{mT}%
$ \cite{nonlinear dynamics}.

In summary, we have succeeded to tune our superconducting microwave resonators
into critical coupling condition using input power, ambient temperature and
applied magnetic field. By data fitting we have been able to extract
quantitatively the coupling parameters $\gamma,g_{1},g_{2}$ and identify the
dominant factor responsible for the coupling change in each case. In addition
we have briefly discussed some of the possible applications of this variable
coupling mechanism in exhibiting some important quantum phenomena in the
microwave regime.

E.B. would especially like to thank Michael L. Roukes for supporting the early
stage of this research and for many helpful conversations and invaluable
suggestions. Very helpful conversations with Bernard Yurke are also gratefully
acknowledged. This work was supported by the German Israel Foundation under
grant 1-2038.1114.07, the Israel Science Foundation under grant 1380021, the
Deborah Foundation and Poznanski Foundation.

\bibliographystyle{plain}
\bibliography{apssamp}
\enlargethispage{-5in}%

\end{document}